\documentclass[aip,
 reprint,
 amsmath,amssymb,
]{revtex4-2}

\usepackage{graphicx}
\usepackage{dcolumn}
\usepackage{bm}

\usepackage[utf8]{inputenc}
\usepackage[T1]{fontenc}
\usepackage{mathptmx}
\usepackage{etoolbox}

\usepackage{chemformula}
\usepackage{chemfig}
\usepackage{mhchem}
\usepackage{physics} 
\usepackage{amsfonts}

\makeatletter
\def\@email#1#2{%
 \endgroup
 \patchcmd{\titleblock@produce}
  {\frontmatter@RRAPformat}
  {\frontmatter@RRAPformat{\produce@RRAP{*#1\href{mailto:#2}{#2}}}\frontmatter@RRAPformat}
  {}{}
}%
\makeatother
\begin{document}


\title{ Molecular dynamics-driven global tetra-atomic potential energy surfaces: Application to the AlF dimer}

\author{Xiangyue Liu}%
\author{Weiqi Wang}

\affiliation{Fritz-Haber-Institut der Max-Planck-Gesellschaft, Faradayweg 4-6, 14195 Berlin, Germany
}%

\author{Jes\'us P\'erez-R\'ios}
\affiliation{Department of Physics and Astronomy, Stony Brook University, Stony Brook 11794, New York (USA)}
\affiliation{Institute for Advanced Computational Science, Stony Brook University, Stony Brook, NY 11794-3800, USA}%
\email{jesus.perezrios@stonybrook.edu}

\date{\today}

\begin{abstract}

In this work, we present a general machine learning approach for full-dimensional potential energy surfaces for tetra-atomic systems. Our method employs an active learning scheme trained on {\it ab initio} points, which size grows based on the accuracy required. The training points are selected based on molecular dynamics simulations, choosing the most suitable configurations for different collision energy and mapping the most relevant part of the potential energy landscape of the system. The present approach does not require long-range information and is entirely general. As an example, we provide the full-dimensional AlF-AlF potential energy surface, requiring $\lesssim 0.1\%$ of the configurations to be calculated {\it ab initio}. Furthermore, we analyze the general properties of the AlF-AlF system, finding key difference with other reported results on CaF or bi-alkali dimers.

\end{abstract}

\maketitle

\begin{quotation}

\end{quotation}

\section{Introduction}

The concept of potential energy surface (PES) is critical for understanding atomic and molecular systems' spectroscopic and scattering properties. As a result, the development of accurate potential energy surfaces (PESs) is one of the most active research areas in chemical physics. Using machine learning (ML) methods, new techniques have been developed for constructing accurate PESs. Precisely, the system's energy is calculated via high-level electronic structure methods for a given set of configurations used to train and test a given ML algorithm. Next, the same ML model is employed to predict the energy of new configurations. Several ML methods have been proposed ~\cite{ML1,ML2,ML3,ML4,christianen2019six,Activelearning,Activelarning2}, for example, one pioneering approach is the permutation invariant polynomial-neural network (PIP-NN)~\cite{jiang2013permutation}, in which permutation invariant polynomials~\cite{braams2009permutationally} depending on the inverse of interatomic distances or Morse-like potentials~\cite{permutation,permutation1,permutation2,permutation3,permutation4}, for a given geometry, are used as the input of a feed-forward neural network, yielding promising results up to systems with 15 degrees of freedom~\cite{Bin2016}. However, for tetra-atomic systems, it requires more than 70000 \textit{ab initio} points to reach a global root mean square error of 40~cm$^{-1}$ ~\cite{Bin2016}. In a different vein, it has been possible to use piece-wise Ml methods such that a Gaussian process regression (GPR) is employed in the short-range interaction region. At the same time, the long-range tail of the potential is fitted given a particular functional form~\cite{christianen2019six}. In this approach, the overall accuracy is not as good as in the case of PIP-NN. However, with only 2000 \textit{ab initio} points, it was possible to reach a root mean square error between 140 and 4~cm$^{-1}$, depending on the region of the PES. Therefore, using more training points in a particular region, the accuracy of the PES improves, as it has been shown via active learning approaches~\cite{Activelearning,Activelarning2,rasheeda2022high}.

Most ML models rely on uniform sampling of the degrees of freedom or interatomic distances. Hence, many training points are required to achieve a given accuracy. However, when calculating scattering properties based on a given underlying PES, it is well-known that the collision energy (or temperature in the case of a canonical ensemble) establishes the relevance of specific configurations, generally related to transition states and local minima. Therefore, when ML-based PESs are used for dynamics calculations, the prediction for relevant configurations may have considerable uncertainty, thus, yielding significant errors in the scattering observables. However, this situation can be avoided after identifying the most pertinent configurations contributing to a given collision energy and including those into the training set, reducing the number of required \textit{ab initio} points while reaching a high accuracy. 

This paper presents a general method to construct PESs for tetra-atomic systems based on an ML approach exploiting a novel selection of atomic configurations. The training configurations are chosen from molecular dynamics (MD) simulations in various temperatures to ensure they sample the same configurational space as their practical usage. Next, the training set is enlarged by including more required configurations via an active learning scheme, leading to an optimal strategy for dynamics calculations for tetra-atomic systems. As a proof of concept, we apply our method to the PES of the AlF dimer, being of interest in the study of ultracold molecules from direct cooling techniques~\cite{Truppe2019, Simon, Max} and showing very intriguing chemistry in buffer gas cells~\cite{Liu2022,Sidney2022}. 




\section{Theory and methods{\label{sec:method}}}

In molecular interactions, the Coulomb and exchange interactions dominate at short range. In contrast, the dispersion and induction interactions dominate at distances beyond the LeRoy radius. As a result, a machine learning-based PES should encode the system's geometry with sufficient structural information to establish the structure-energy relationship as a non-linear regression problem.

\subsection{Dimer representation}

\begin{figure}[ht]
        \centering
        \includegraphics[width=1\linewidth]{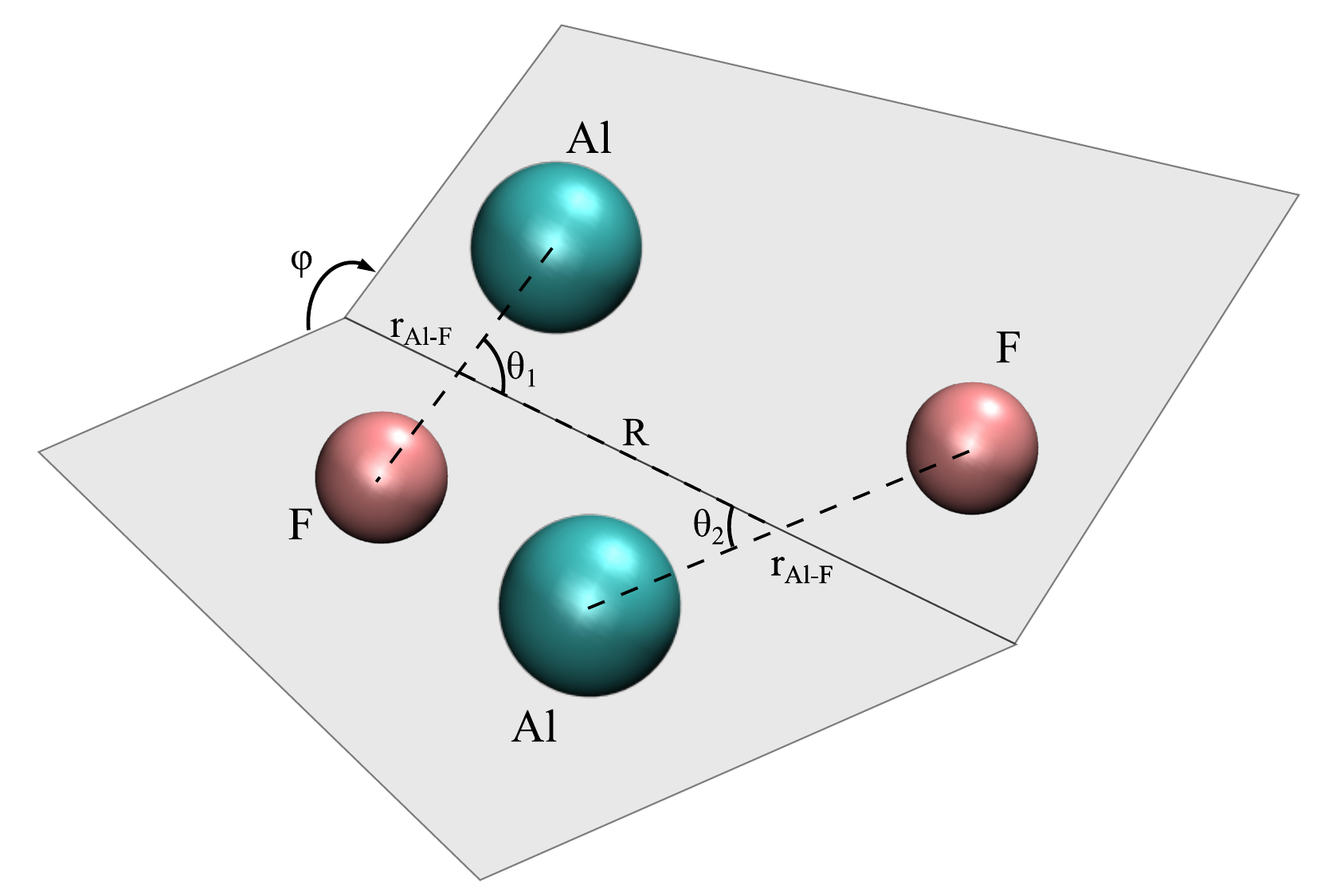}
        \caption{Jacobi coordinates for the AlF dimer. The Jacobi coordinates consist of two interatomic distances $r_{\text{Al-F}}$, two azimuthal angles $\theta_1$ and $\theta_2$, one torsion angle $\phi$, and the inter-molecular distance $R$ associated to the vector joining the center of mass of the two molecules.}
        \label{fig:Jacobi}
    \end{figure}

The interaction between two diatomic molecules has six degrees of freedom, and it is better described using Jacobi coordinates. As shown in Fig.\ref{fig:Jacobi} for the case of the AlF-AlF dimer, the Jacobi coordinates consist of two interatomic distances, three angles, and the intermolecular distance $R$ joining the center of mass of the two molecules. In principle, a representation should obey the same symmetries preserving or changing the energy \cite{bowman2011high,jiang2013permutation,mbtrhuo2017unified}. In this respect, Jacobi coordinates maintain translational and rotational symmetry. In contrast, they can not keep the permutational invariant of the system. However, the permutational invariant symmetry is restored using a representation given by a series of $n$-body interaction terms $\{G(\mathbf{x},\mathbf{i})\}$ with a symmetrization operator $\hat{S}$ accounting for the permutational symmetry of the system.

The energy for a given configuration of atoms is described via a Gaussian process as

\begin{equation}
    E(\mathbf{x}) \sim \text{GP}(m(f(\mathbf{x})), k(f(\mathbf{x}), f(\mathbf{x}'))),
\end{equation}
where
\begin{equation}
f(\mathbf{x}) = \hat{S} \{ G(\mathbf{x},\mathbf{i}) \},
\end{equation}
$\mathbf{x}$ represents any geometry variable, and $i$ labels atoms within the molecular system. In order to capture both Coulomb and long-range interactions, we consider two-body interactions described by the inverse interatomic distances $1/\bar{r}_{ij}$ between atoms $i$ and $j$, as well as Morse-like exponential terms $e^{-\bar{r}_{ij}}$.~\footnote{We have also tested the performance with additional representations of higher orders, e.g. three-body polar angles and four-body dihedral angles, however, the improvement is negligible in the AlF-AlF system.} To differentiate the chemical distinctions between various pairs of atoms, the interatomic distance $\bar{r}_{ij}$ has been normalized by the equilibrium interatomic distance $r^*_{ij}$, i.e. $\bar{r}_{ij}=r_{ij}/r^*_{ij}$.

\subsection{\textit{Ab initio} calculations}

We have calculated \textit{ab initio} energies using coupled-cluster theory with single, double, and perturbation triples [CCSD(T)] implemented in the Molpro package~\cite{molpro,werner2012molpro}. During MD simulations, the \textit{ab initio} forces have been calculated at the second‐order Møller–Plesset perturbation theory (MP2) level. The calculations were performed with the aug-cc-pVQZ basis set~\cite{avqzdunning1989gaussian,avqzkendall1992electron, avqzwoon1993gaussian}. 

\subsection{Construction of the PES}
\label{Construction_PES}

Herein, GPR \cite{gprwilliams2006gaussian} is used to find the relationship between the structural representation and the energy of the system, leading to a non-parametric fitting of the PES, i.e., no functional form is assumed for the representation of the PES. GPR assumes a Gaussian distribution of functions over a function space, and kernel functions are employed as the covariance of the Gaussian process random variables. Once exposed to a data set, a posterior of the function distribution for new structures can be obtained following the Bayes theorem.



The kernel function determines the shape of the prior and posterior distribution. We have tested different kernel function combinations by analyzing the mean absolute error (MAE) and median absolute error of the test set predictions. As a result, the Matérn kernel with $\nu=5/2$ turns out to be the most accurate kernel. Furthermore, the non-fixed hyperparameters of the kernels are optimized by maximizing the marginal likelihood with gradient ascent.   

The reference datasets have been created using initial sets of energies from eight \textit{ab initio} MD trajectories and later extended configurations obtained through active learning during the MD simulation. The initial MD trajectories were run within the canonical ensemble at 200 and 800~K, starting from different initial configurations. As a result, we sample configurations relevant to short-range as well as long-range interactions. Specifically, the sampled data's intermolecular distance $R$ ranges from approximately 1.5 to 17.5 \AA ~. From these trajectories, different numbers of configurations are randomly chosen and taken as training sets. In contrast, the GPR predictions are tested against the configurations visited in a different MD trajectory of 3633 steps. In a second version of the approach, we have selected 2271 and 5026 ``landmarks'' for two additional training sets, which are the configurations with the longest high-dimensional distances from the others in the representation space, chosen from the eight trajectories and tested against the same MD trajectory of 3633 steps.


    \begin{figure}[h!]
        \centering
        \includegraphics[width=1\linewidth]{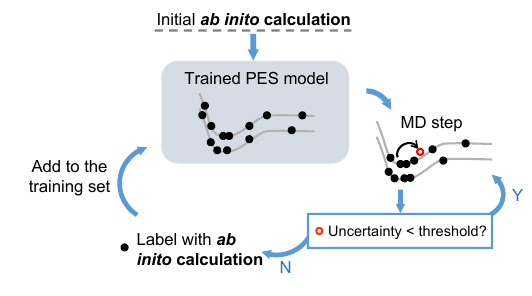}
        \caption{Schematic diagram of the active learning approach. Starting from an initial trained model, a MD simulation is performed with the force calculated by the finite difference of energies predicted by the model. If the energy prediction uncertainty of the new MD step is larger than the threshold, then the configuration will be calculated with \textit{ab initio} method and added to the training set. }
        \label{fig:active_learning}
    \end{figure}

\subsection{Active learning scheme}
\label{sec.activelearning}

It is well-known that the accuracy of ML or any fitting approach depends on the region of the PES under consideration. Particularly, the repulsive wall of the PES is the hardest to describe, leading to considerable uncertainty in the predictions. In the meantime, some regions in the training sets might be sparsely sampled, leading to larger prediction uncertainty in those areas. However, these effects can be mitigated by including more configurations in the training set. But, more configurations require more \textit{ab initio} points, leading to longer computational times. Here, we avoid this situation via an active learning approach.

Our active learning scheme is illustrated in Fig.~\ref{fig:active_learning}. First, we are ready to predict the energy value of a new configuration after training the model on a given initial training set of \textit{ab initio} points. We will have a prediction based on the GPR and its uncertainty for this new configuration sampled by the next MD step. Then, if the uncertainty is smaller than a given threshold, we use the GPR prediction. Otherwise, we require an \textit{ab initio} point. In this case, the new configuration is included in the training set for the next prediction and will be used in the next MD step. This process is repeated as many times as required until the configuration sampling converges. 




\section{Results and discussions}

\begin{figure}[ht]
        \centering
        \includegraphics[width=0.8\linewidth]{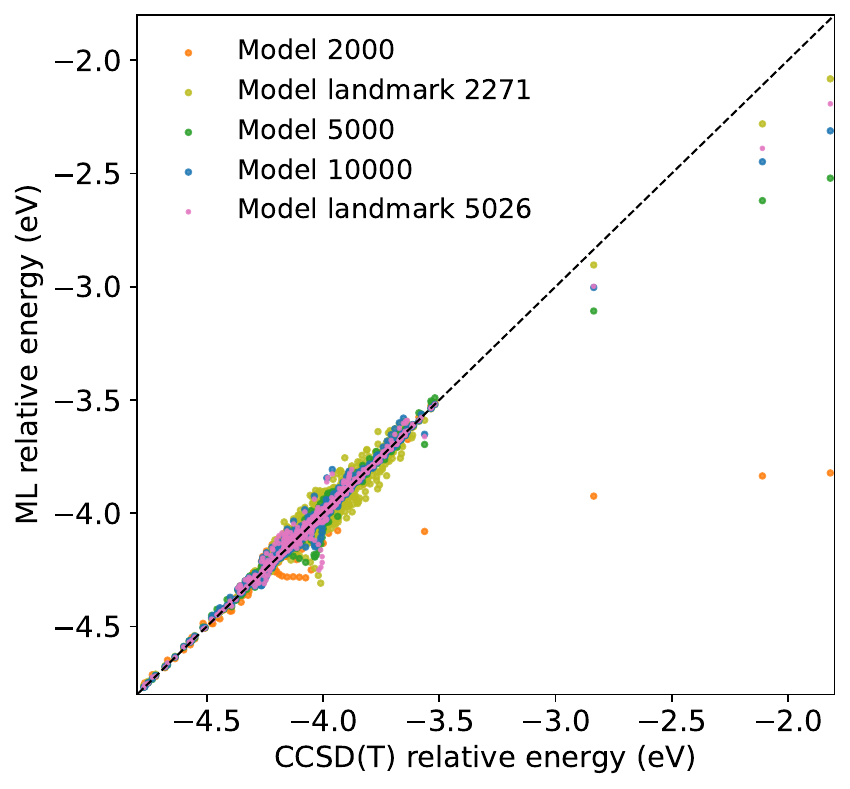}
        \caption{The GPR-predicted energy versus the CCSD(T)/aug-cc-pVQZ energy, shown for different initial training sets.}
        \label{fig:energy_ML_vs_CCSDt}
    \end{figure}

To test the accuracy and efficiency of the present method, the PES models have been trained with several different initial training sets constructed as described in Sec.\ref{Construction_PES}. Fig.\ref{fig:energy_ML_vs_CCSDt} compares the model predictions against a test set generated from an \textit{ab initio} MD trajectory of 3633 steps simulated at 800~K. As expected, the performance of the models trained with randomly-selected configurations improves with increasing training set size. Increasing the size of the training set from 2000 to 5000 reduces the error by approximately 25$\%$, as shown in Table~\ref{tab:error_range}, with the mean and median absolute errors for the 5000-configuration training set of 0.78 meV/atom and 0.018 meV/atom, respectively, demonstrating high accuracy of the models with more training data. With 10000 training configurations, the accuracy remains the same as with 5000 configurations, but the predictions on the outliers with the highest energies become closer to the reference energy. On the other hand, despite the impressive overall accuracy of the models, they show a lower accuracy for configurations yielding high interaction energies, corresponding to the repulsive region with short intermolecular distance $R$.

\begin{table*}[]
\footnotesize
\caption{The mean absolute error and median absolute error for the test set with 3633 MD steps, presented in meV/atom, with the errors reported for the entire test set, as well as for configurations within different intermolecular ranges of $R$ (in \AA).}
\label{tab:error_range}
\begin{tabular}{lclllllllllllllll}
\hline
              & \multicolumn{8}{c}{Mean absolute error (meV/atom)}                                                            & \multicolumn{8}{c}{Median absolute error (meV/atom)}                                                           \\

              & all & {[}0,2.5{]} & {[}2.5,5{]} & {[}5,7.5{]} & {[}7.5,10{]} & {[}10,12.5{]} & {[}12.5,15{]} & {[}15,17.5{]} & all & {[}0,2.5{]} & {[}2.5,5{]} & {[}5,7.5{]} & {[}7.5,10{]} & {[}10,12.5{]} & {[}12.5,15{]} & {[}15,17.5{]} \\
              
              \hline
Model 2000    & 1.18  & 15.1        & 2.9         & 0.41        & 0.088        & 0.013         & 0.0088        & 0.022         & 0.024 & 4.49        & 1.51        & 0.25        & 0.048        & 0.009         & 0.0068        & 0.0069        \\
Model 5000    & 0.78  & 6.80        & 2.67        & 0.36        & 0.070        & 0.014         & 0.0077        & 0.0080        & 0.018 & 3.05        & 1.37        & 0.19        & 0.031        & 0.010         & 0.0055        & 0.0047        \\
Model 10000   & 0.85  & 6.50        & 3.06        & 0.49        & 0.071        & 0.014         & 0.0077        & 0.0071        & 0.019 & 2.74        & 1.65        & 0.19        & 0.071        & 0.010         & 0.0052        & 0.0039        \\
Landmark 2271 & 0.81  & 4.70        & 3.16        & 0.52        & 0.074        & 0.032         & 0.023         & 0.027         & 0.044 & 2.24        & 1.57        & 0.27        & 0.047        & 0.021         & 0.019         & 0.021         \\
Landmark 5026 & 0.85  & 5.99        & 3.18        & 0.51        & 0.065        & 0.016         & 0.012         & 0.012         & 0.024 & 2.14        & 1.63        & 0.22        & 0.034        & 0.010         & 0.0093        & 0.0089       \\

\hline

\end{tabular}
\end{table*}

\begin{figure}[ht]
        \centering
        \includegraphics[width=1\linewidth]{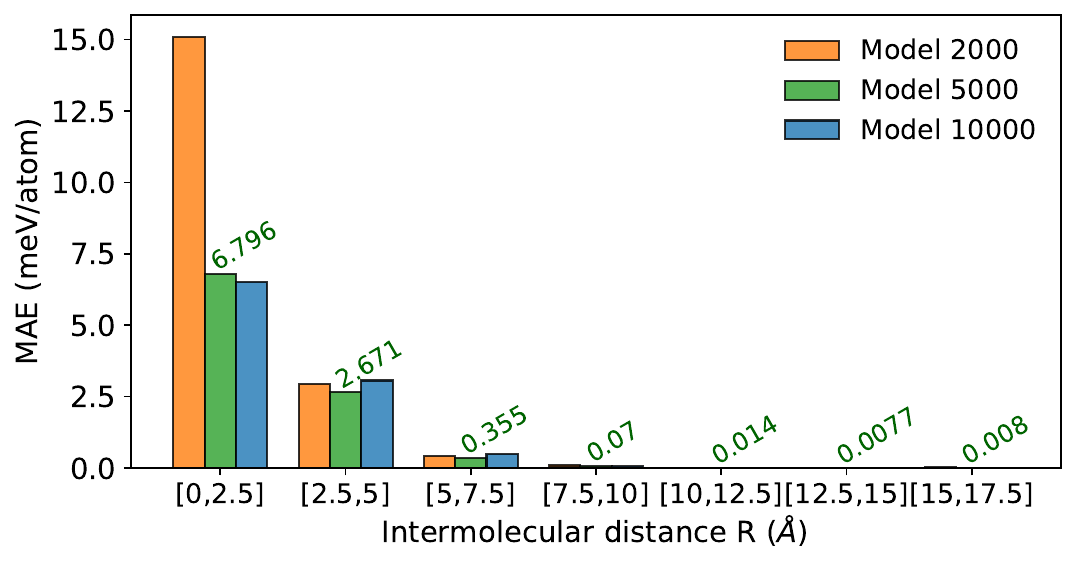}
        \caption{MAE in different regions of intermolecular distance $R$, tested on the test set with 3633 MD steps. The MAE from the model with 5000 training data is labeled in green.}
        \label{fig:MAE_vs_R}
    \end{figure}

To further understand the accuracy of our model as a function of $R$, we have computed the mean absolute error for different regions of $R$, as shown in Fig.~\ref{fig:MAE_vs_R}. This figure shows that the region with $R \in [0,2.5]$, shows the largest mean absolute error across all models. On the contrary, for $R > 7.5$ \AA\, the errors are less than 0.1 meV/atom with all models. A more detailed study is presented in Table~\ref{tab:error_range}, where we report on the mean and median absolute errors, and both error estimators point to a lack of configurations with $R \in [0,2.5]$.

As introduced in Sec.~\ref{Construction_PES}, two additional training sets have been constructed with 2271 and 5026 landmark configurations. This approach aims at reducing the number of training data, thereby expediting the training process, assuming that the landmarks are representative of the sampled configurational space. Indeed, the training set with landmarks provides similar overall accuracy as the models constructed from the training set with 10000 randomly selected configurations. However, keeping in mind that the accuracy of randomly sampled points is not enough in the short-range region, and given that trajectories in the thermal regime will visit the short-range interaction region, we conclude that using the most relevant configurations visited by the MD simulations as training points leads to a more accurate PES model.  




\begin{figure}[ht]
        \centering
        \includegraphics[width=0.8\linewidth]{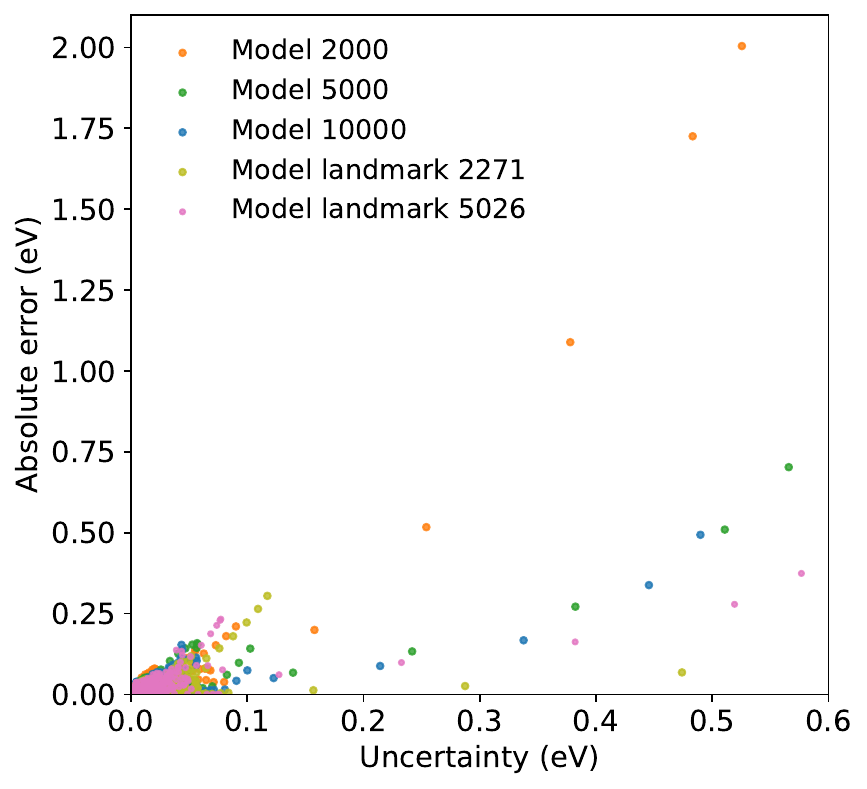}
        \caption{The relationship between the absolute error and the uncertainty of GPR prediction.}
        \label{fig:error_vs_uncertainty}
    \end{figure}

We switch to our active learning scheme introduced in Sec.~\ref{sec.activelearning} to generate more training points. First, we check the correlation between the prediction uncertainty and the absolute error across all models. The results are shown in Fig.~\ref{fig:error_vs_uncertainty}, where a correlation is noticed between the absolute error of a GPR prediction and its uncertainty. Hence, the uncertainty of the prediction is a good parameter to estimate the quality of the prediction when compared against a given accuracy threshold.  

\begin{figure}[ht]
        \centering
        \includegraphics[width=0.8\linewidth]{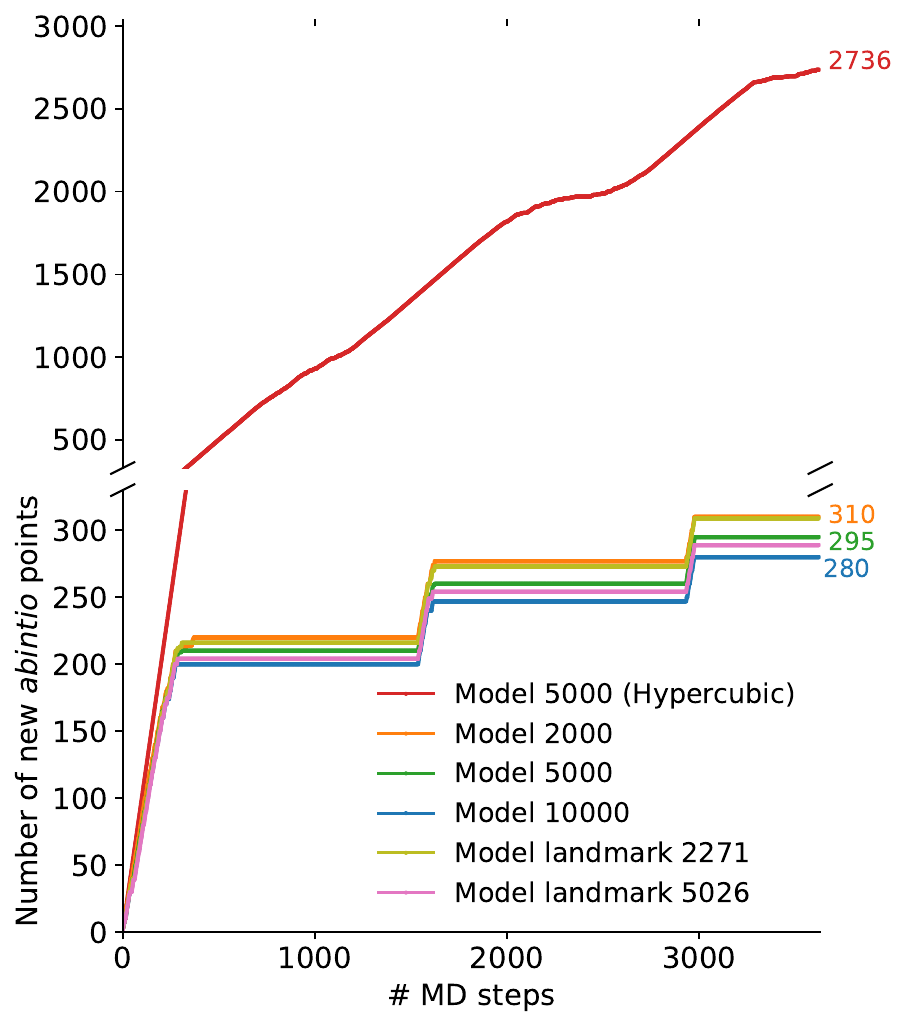}
        \caption{Number of new \textit{ab initio} points additionally required in active-learning during the MD simulation, tested on the MD trajectory with 3633 steps. The uncertainty $>0.01$ eV has been used as the criterion for additional \textit{ab initio} calculations.}
        \label{fig:N_new_points}
    \end{figure}

To compare the efficiency of the initial PES models trained using training sets of varying sizes, we have applied our active learning strategy based on different initial training sets, and the results are displayed in Fig.~\ref{fig:N_new_points}. All models are exposed to 3633 new configurations from a single MD simulation. With the uncertainty threshold set to be 0.01~eV, the models trained on the configurations randomly selected from eight MD trajectories require around $10\%$ of the configurations to be calculated \textit{ab initio}. As expected, the models with more training data require fewer additional \textit{ab initio} calculations. We have also observed a convergence behavior of new \textit{ab initio} points with respect to the number of MD steps. This indicates that in longer simulations, it is likely that fewer than $10\%$ of the configurations need to undergo \textit{ab initio} calculations. Similarly, training sets with landmark configurations show an almost identical efficiency. On the contrary, the initial model trained on hypercubic grids requires the \textit{ab initio} evaluation of almost $65\%$ of the configurations, showing the poor efficiency of Latin hypercube sampling schemes specifically in the case of MD simulations. Therefore, a Latin hypercube sampling will lead to long computational times, whereas a sampling using our MD-driven PES model will be more efficient.

\begin{figure}[ht]
        \centering
        \includegraphics[width=0.75\linewidth]{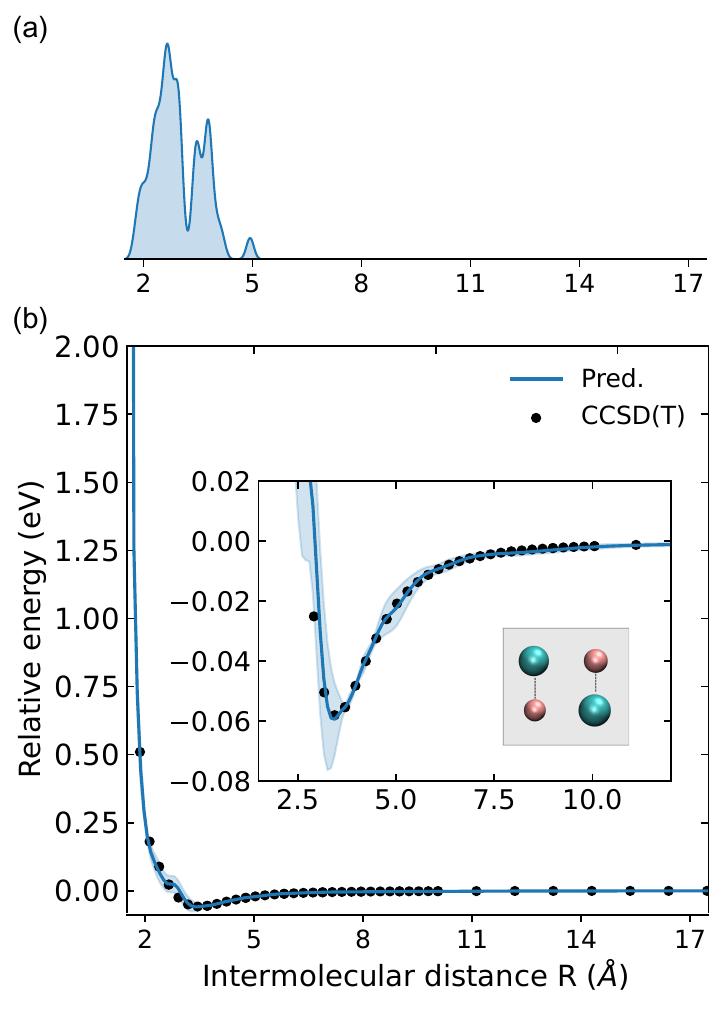}
        \caption{(a) Distribution of intermolecular distance $R$ of the configurations requiring additional \textit{ab initio} calculations in the MD simulation of AlF-AlF complex.
        (b) An one-dimensional profile of the PES, showing the \textit{ab initio} and predicted energies with the prediction uncertainty indicated by shaded regions. The model is trained with a training set of 24,403 configurations. The total energy is referenced to the energy of dissociated AlF-AlF complex.}
        \label{fig:distribution_R_PEC}
    \end{figure}

In practice, MD simulations can be much longer than this test case to sample the configurational space and reach equilibrium adequately. To further assess the feasibility of our approach in realistic simulations, we have performed a simulation using the replica-exchange molecular dynamics (REMD) method \cite{REMDmarinari1992simulated, REMDsugita1999replica}. REMD is an enhanced sampling technique that facilitates the attainment of dissociation equilibrium at low temperatures more efficiently. In particular, we simultaneously simulated 10 trajectories at temperatures ranging from 200 K to 1000 K. The total simulation time reaches 5.4 ns in each of the ten replicas. The active learning approach described above has been employed with an initial PES model trained with 22,365 configurations obtained from all of the nine MD trajectories. During the REMD simulation, a total of 2038 configurations ($\sim0.008\%$) out of 26,891,350 REMD steps have been selected to be further calculated with CCSD(T) method, based on a criterion of prediction uncertainty $0.05$ eV. These additional configurations are primarily located at the regions with intermolecular distance $R<5$ \AA, as depicted in Fig.~\ref{fig:distribution_R_PEC} (a). The accuracy of this model is tested by a comparison between the CCSD(T) reference energies and the model predictions for a one-dimensional profile of the PES, as shown in Fig.~\ref{fig:distribution_R_PEC} (b). The model demonstrates a high global accuracy, although the uncertainty is notably more significant in the repulsive region than in the long-range one.

\section{Properties of the PES}

We start by analyzing the general properties of the \textit{ab initio} PES, and the results are summarized in Table~\ref{tab:energies}. Among them,  it is worth highlighting that the reactants need to surmount 11.56~eV of energy to yield \ce{AlF} + \ce{AlF} $\rightarrow$ \ce{Al2} +  \ce{F2}. Therefore, this reaction is highly endothermic.  


\begin{table}[h]
\caption{The CCSD(T)/aug-cc-pVQZ relative energies of various configurations. The energies are referenced to the potential energy of two AlF molecules with the intermolecular distance $R$ = 20 \AA. In this case, the geometry of AlF molecule has been fixed to its CCSD(T)-optimized geometry with $d$(AlF) = 1.669 \AA.}
\label{tab:energies}
\begin{tabular}{cc}
\hline
Configuration                            & Relative energy (eV) \\ 
\hline
AlF + AlF ($d$(AlF-AlF) = 20 \AA)        & 0.0                  \\ 
AlF-AlF complex                          & -0.696               \\ 
Al$_2$ + F$_2$ ($d$(Al$_2$-F$_2$) = 20 \AA)    & 11.561               \\ 
Dissociated 4 atoms ($d$(Al-F) = 20 \AA) & 18.167               \\ 
\hline
\end{tabular}
\end{table}

The stable configuration of the AlF-AlF complex has been optimized with the PES model with the Fast Inertial Relaxation Engine (FIRE) method \cite{bitzek2006structural}. The resulting geometry shows a D$_{2h}$ symmetry, with all Al-F bond lengths being the same. This geometry reproduces CCSD(T)-optimized geometry, with differences in Al-F bond lengths of less than 0.0004 \AA, while the predicted energy is 0.0025 eV higher than CCSD(T) result. It turns out that there is no barrier for the atom-exchange reaction Al$^{(1)}$F$^{(2)}$ + Al$^{(3)}$F$^{(4)}$ = Al$^{(1)}$F$^{(4)}$ + Al$^{(3)}$F$^{(2)}$. Indeed, this reaction has frequently been observed during the reported MD simulations for the AlF dimer. 


    \begin{figure}[h!]
        \centering
        \includegraphics[width=0.5\linewidth]{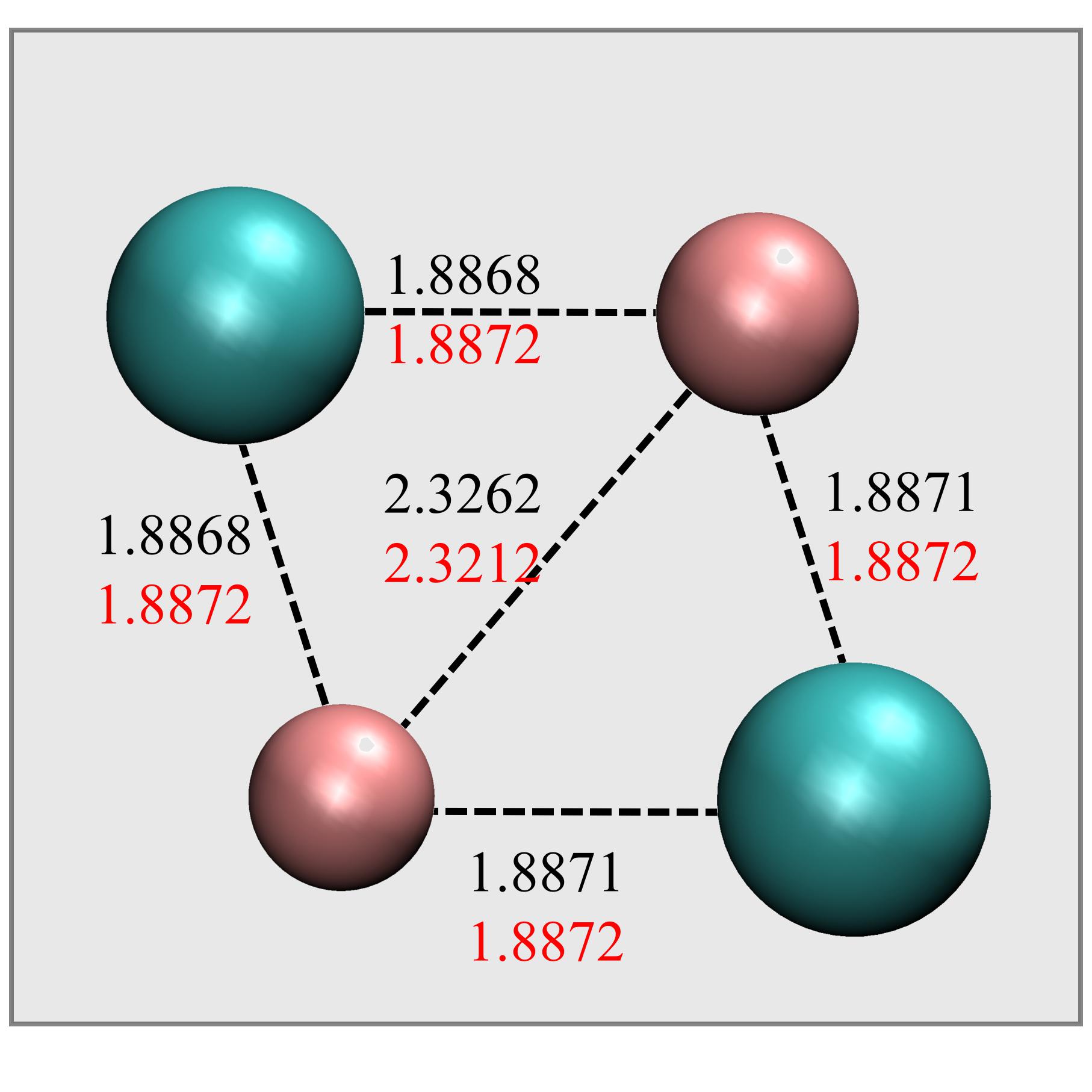}
        \caption{Configurations of AlF-AlF complex, optimized from the ML-fitted PES (black), and compared with CCSD(T)/aug-cc-pVQZ optimized geometry (red). The ML-predicted relative energy of the complex is -0.695 eV, referenced to the potential energy of two AlF molecules with the intermolecular distance $R$ = 20 \AA. }
        
        \label{fig:distribution_R}
    \end{figure}

Interestingly, the D$_{2h}$ configuration is the only stable configuration in the PES, in contrast to the behavior observed in CaF\cite{sardar2023four} or NaK \cite{christianen2019six}. Regardless of the initial configurations used, the geometry optimization of AlF-AlF consistently converges to the stable D$_{2h}$ configuration. Other configurations, such as the local minimum configuration found for the CaF-CaF complex with C$_s$ symmetry, are unstable for the AlF-AlF complex. 

\section{Conclusions}

In this work, we have introduced a general ML method to fit tetra-atomic potential energy surfaces. The method exploits the most relevant configurations for a given MD simulation at a given temperature as the training set to feed a GPR to predict the energy of new configurations. This process can be adapted to satisfy a given threshold criteria, reaching the required accuracy for a given system. As a result, it is possible to construct highly accurate tetra-atomic potential energy surfaces requiring $\lesssim 0.1\%$ of the configurations to be calculated \textit{ab initio}. Hence, delivering accurate results with low computational effort. 

We have applied the method to AlF-AlF as proof of concept, finding a single minimum contrary to previous results in bi-alkali molecules or CaF. The four-body complex shows a binding energy of 0.69~eV. Therefore, it may be the case that AlF-AlF offers a very different complex lifetime than CaF or bi-alkali systems. This topic is under consideration in our group and will be published elsewhere. 

Finally, our method opens up a new avenue for MD simulations in tetra-atomic systems by delivering accurate but computationally cheap, full-dimensional potential energy surfaces. On the other hand, extending our approach to a larger system is possible when combined with a representation that considers the system's symmetry.

\begin{acknowledgments}
J. P.-R. acknowledges discussions with Gerrit Groenenboom and the support of the Air Force Office of Scientific Research under award number FA9550-23-1-0202. 
X. Liu acknowledges the support of the Deutsche Forschungsgemeinschaft (DFG – German Research Foundation) under grant number PE 3477/2 - 493725479.
W. W acknowledges funding by the Max Planck-Radboud University Center for Infrared Free Electron Laser Spectroscopy. 


\end{acknowledgments}

\section*{Data Availability Statement}
The datasets and the computer codes for the training and test of the potential energy surface model are available at the git repository \url{https://github.com/onewhich/AlF_dimer}.



\nocite{*}
\bibliography{apssamp}

\end{document}